# Three-way Mixed Effect ANOVA to Estimate MRMC Limits of Agreement

Si Wen[a]* and Brandon D. Gallas[a]

[a] *CDRH/OSEL Division of Imaging, Diagnostics, and Software Reliability, U.S. FDA, Silver Spring, USA*

* Correspondence to: Si Wen. Email: si.wen@fda.hhs.gov

# Three-way Mixed Effect ANOVA to Estimate MRMC Limits of Agreement

**Abstract**: When evaluating the clinical performance of a medical imaging device, a multi-reader multi-case (MRMC) analysis is usually applied to account for both case and reader variability. For a clinical task that equates to a quantitative measurement, an agreement analysis such as a limits of agreement (LOA) method can be used to compare different measurement methods. In this work, we introduce four types of comparisons; these types differ depending on whether the measurements are within or between readers and within or between modalities. A three-way mixed effect ANOVA model is applied to estimate the variances of individual differences, which is an essential step for estimating LOA. To verify the estimates of LOA, we propose a hierarchical model to simulate quantitative MRMC data. Two simulation studies were conducted to validate both the simulation and the LOA variance estimates. From the simulation results, we can conclude that our estimate of variance is unbiased, and the uncertainty of the estimation drops as the number of readers and cases increases and rises as the value of true variance increases.

## 1. Introduction

Multi-reader multi-case (MRMC) analysis is commonly utilized when evaluating the clinical performance of a medical imaging device. (Wagner, Metz, & Campbell, 2007; Gallas, et al., 2012) In an MRMC study, a set of clinical readers (e.g., radiologists or pathologists) evaluates images from a set of patient cases for a specific clinical task under two reading conditions or modalities. One modality is usually the new technology, the test modality. The other modality is the reference technology, which may be current clinical practice. We compare the reader-averaged performance of the test and reference modalities to see if the new technology is as good or better than the reference.

For some of the clinical tasks, there is binary ground truth or an ordinal reference standard associated with the clinicians' evaluation. For example, the clinicians' decision about whether a patient should be recalled or not is compared against the binary cancer status of the patient (Gallas, et al., 2019) and the abnormality detection from a medical image is compared against the binary biopsy gold standard (Hillis, Obuchowski, Schartz, & Berbaum, 2005). In these situations, the MRMC receiver operating characteristic (ROC) analysis (Hillis, Obuchowski, Schartz, & Berbaum, 2005; Obuchowski N. A., 1995; Dorfman, Berbaum, & Metz, 1992; Barrett, Kupinski, & Clarkson, 2005; Gallas B. D., 2006; Gallas, Bandos, Samuelson, & Wagner, 2009) is often used for comparing the reader-averaged performance.

For some other clinical tasks, there is no binary ground truth for each case and the reference standard is a quantitative measurement. Examples include the number of mitotic figures in a glass pathology slide as seen with digital whole slides imaging versus a microscope (Tabata, et al., 2019) and the acetabular version angle as seen with magnetic resonance imaging (MRI) versus computed tomography (CT) (Obuchowski, Subhas, & Schoenhagen, 2014). We refer to such studies as agreement studies (Barnhart, Haber, & Lin, 2007; Gallas, Anam, Chen, Wunderlich, & Zhang, 2016), studies in which the "closeness" between measurements from different methods or readers is assessed.

One of the widely used statistical methods for assessing quantitative agreement in medical literature is the limits of agreement (LOA) method (Bland & Altman, 1986; Bland & Altman, 1999). The main idea of the LOA method is to generate the $(1-\alpha)\%$ probability interval that covers the middle $1-\alpha$ probability of the distribution of differences in observations between two readers or modalities. Let $D_k = Y_{1k} - Y_{2k}$ be the difference in measurements from two different readers or two different modalities for case $k (k = 1, \dots, K)$, then the 95% LOA are $\mu_D \pm 1.96\sigma_D$, where $\mu_D$ and $\sigma_D$ are the mean and standard deviation of the differences. LOA are similar to but different from the confidence interval for the mean difference, $\mu_D \pm 1.96\, \sigma_D/\sqrt{K}$.

The latter one will be narrower as the sample size increases; it reflects the uncertainty in the **mean** difference instead of the uncertainty of the differences themselves. The estimated LOA are often presented as two horizontal lines in the Bland-Altman plot as shown in Figure 1, in which the individual differences are shown with the difference $D_k$ as y-value and the average of the measurements $(Y_{1k} + Y_{2k})/2$ as x-value.

In this paper, we will introduce four types of LOA for an MRMC agreement study; these types differ depending on whether the measurements are within or between readers and within or between modalities. The standard methods need to be generalized to account for the fact that MRMC data is not independent and identically distributed. The data are likely correlated when the readers evaluate the same cases and within- and between- reader agreement are expected to depend on the specific readers (their background, training, and expertise). We are expanding on related work that has generalized the LOA method to treat multiple readers (Jones, Dobson, & O'Brian, 2011; Christensen, Borgbjerg, & Børty). We are also providing the theoretical foundation for previously published within-reader agreement results (Tabata, et al., 2019). We will illustrate how to apply a three-way mixed effect ANOVA model to estimate the variance of the differences. To validate our estimation method, we developed a hierarchical simulation model for quantitative MRMC data, and we will discuss why an additive linear simulation model is not adequate. Two simulation studies were conducted to validate both the simulation and the LOA variance estimates.

## 2. Method

### *2.1 Limits of agreement for MRMC study*

Suppose $X_{ijk}$ denotes the measurement for case $k$ $(k = 1, \dots, K)$ from reader $j$ $(j = 1, \dots, J)$

under modality $i$ ($i = 1,2$) in an MRMC study comparing two different modalities, where $i = 1$ and $i = 2$ indicate test modality and reference modality respectively. We focus on a fully-crossed study in which each reader provides a measurement for each case for both modalities. Let $D_{jj',k}^{ii'} = X_{ijk} - X_{i'j'k}$ denote the difference between two measurements and $\bar{D}_{jj'}^{ii'}$ denote the difference averaged across all the cases.

There are four types of case-specific and case-averaged differences:

(1) Within-Reader Within-Modality (WRWM) differences, $D_{jj,k}^{ii} = X_{ijkl} - X_{ijkl'}$: The two measurements are from the same reader $j$ using the same modality $i$. To calculate this difference, replicate readings from all the readers are required. In the equation, we use subscript $l$ and $l'$ to denote different replicate readings. The case-averaged differences $\bar{D}_{jj}^{ii} = E\left(D_{jj,k}^{ii}|i,j\right)$ are found on the diagonal in the table on the left in Figure 2. We average these over the readers to obtain $\bar{D}_{WR}^{ii} = E\left(D_{jj,k}^{ii}|i\right)$.

(2) Between-Reader Within-Modality (BRWM) differences, $D_{jj',k}^{ii} = X_{ijk} - X_{ij'k}$: The two measurements are from different readers $j$ and $j'$ using the same modality $i$. The case-averaged differences $\bar{D}_{jj'}^{ii} = E\left(D_{jj',k}^{ii}|i,j,j'\right)$ are found on the off-diagonals in the table on the left in Figure 2. We average these over the pairs of readers to obtain $\bar{D}_{BR}^{ii} = E\left(D_{jj',k}^{ii}|i\right)$.

(3) Within-Reader Between-Modality (WRBM) differences, $D_{jj,k}^{12} = X_{1jk} - X_{2jk}$: The two measurements are from the same reader $j$ using the test modality $i = 1$ and the reference modality $i = 2$. The case-averaged differences $\bar{D}_{jj}^{12} = E\left(D_{jj,k}^{12}|j\right)$ are found on the diagonal in the table on the right in Figure 2. We average these over the readers to obtain $\bar{D}_{WR}^{12} = E\left(D_{jj,k}^{12}\right)$.

(4) Between-Reader Between-Modality (BRBM) differences, $D_{jj',k}^{12} = X_{1jk} - X_{2j'k}$: The two measurements are from different readers $j$ and $j'$ using modalities $i = 1,2$. The case-averaged differences $\bar{D}_{jj'}^{12} = E\left(D_{jj',k}^{12}|j,j'\right)$ are found on the off-diagonals in the table on the right in Figure 2. We average these over the pairs of readers to obtain $\bar{D}_{BR}^{12} = E\left(D_{jj',k}^{12}\right)$.

Given the four types of difference measurements possible in an MRMC study, there are four corresponding types of LOA. Let's take the BRBM difference as an example. The 95% limits of agreement for that group of difference measurements is defined as $\bar{D}_{BR}^{12} \pm 1.96\sqrt{V_{BR}^{12}}$, where $V_{BR}^{12} = Var\left(D_{jj',k}^{12}\right)$ denotes the variance (over readers and cases) of individual BRBM differences. This means that 95% of the BRBM differences would be expected to lie between $\bar{D}_{BR}^{12} - 1.96\sqrt{V_{BR}^{12}}$, and $\bar{D}_{BR}^{12} + 1.96\sqrt{V_{BR}^{12}}$. To construct the four types of limits of agreement for MRMC study, we need to obtain two groups of estimates: estimates of the mean $(\hat{\bar{D}}_{WR}^{ii}, \hat{\bar{D}}_{BR}^{ii}, \hat{\bar{D}}_{WR}^{12}, \hat{\bar{D}}_{BR}^{12})$ and estimates of the variance $(\hat{V}_{WR}^{ii}, \hat{V}_{BR}^{ii}, \hat{V}_{WR}^{12}, \hat{V}_{BR}^{12})$.

### *2.2 Estimates of mean differences*

To estimate the mean differences by a finite sample, we apply the method of moments. The sample mean differences are used to estimate the population mean differences, as follows:

$$\hat{\bar{D}}_{WR}^{ii} = \frac{1}{JK}\sum_j\sum_k\left(X_{ijk1} - X_{ijk2}\right) = \overline{X_{i\cdot\cdot1}} - \overline{X_{i\cdot\cdot2}}\,, \tag{1}$$

$$\widehat{\bar{D}}_{WR}^{12} = \frac{1}{JK}\sum_j \sum_k \left(X_{1jk} - X_{2jk}\right) = \overline{X_{1\cdot\cdot}} - \overline{X_{2\cdot\cdot}}, \tag{2}$$

$$\widehat{\bar{D}}_{BR}^{ii'} = \frac{1}{J^2K}\sum_j \sum_{j'} \sum_k \left(X_{ijk} - X_{i'j'k}\right) = \overline{X_{i\cdot\cdot}} - \overline{X_{i'\cdot\cdot}}. \tag{3}$$

where $\overline{X_{i\cdot\cdot}} = \sum_j \sum_k X_{ijk}/(JK)$ , $i = 1,2$ denotes the sample average measurement across all the readers and cases for a single modality. For the between-modality differences ($i = 1, i' = 2$), the mean estimates for the WRBM and BRBM difference are the same and equal to the difference of average measurements for the two modalities. For the within-modality differences ($i = i' = 1,2$), the mean estimate for the WRWM difference is the difference of mean for two replicates and the one for the BRWM difference is zero.

### *2.3 Using three-way mixed effect ANOVA to estimate the variances of differences*

To estimate the variances of the differences, we build up a three-way mixed effect ANOVA model for the measurement $X_{ijkl}$:

$$X_{ijkl} = \mu + m_i + R_j + C_k + RC_{jk} + mR_{ij} + mC_{ik} + \varepsilon_{ijkl} \tag{4}$$

where $\mu$ denotes the grand mean, the true quantitative value, and $m_i$ represents the fixed effect for modality ($\sum_i m_i = 0$). The other variables are related to the two random effects: reader and case. The variables are independent and normally distributed with mean zero and variances given by $\sigma_R^2$, $\sigma_C^2$, $\sigma_{RC}^2$, $\sigma_{mR}^2$, $\sigma_{mC}^2$, $\sigma_\varepsilon^2$. The mixed effect model we apply here is the unrestricted mixed effect model. We do not force $\sum_i mR_{ij} = 0$, $\sum_i mC_{ik} = 0$ as would be done in a restricted mixed effect model.

From the ANOVA model, the differences can be expressed as the linear combination of the terms in model, and the variances of the differences can be derived as:

$$V_{WR}^{ii} = Var\left(D_{jj,k}^{ii}\right) = Var\left(X_{ijk1} - X_{ijk2}\right) = 2\sigma_{\varepsilon}^{2} \,, \quad (5)$$

$$V_{BR}^{ii} = Var\left(D_{jj',k}^{ii}\right) = Var\left(X_{ijk} - X_{ij'k}\right) = 2\sigma_{R}^{2} + 2\sigma_{RC}^{2} + 2\sigma_{mR}^{2} + 2\sigma_{\varepsilon}^{2} \,, \quad (6)$$

$$V_{WR}^{12} = Var\left(D_{j,k}^{12}\right) = Var\left(X_{1jk} - X_{2jk}\right) = 2\sigma_{mR}^{2} + 2\sigma_{mC}^{2} + 2\sigma_{\varepsilon}^{2} \,, \quad (7)$$

$$V_{BR}^{12} = Var\left(D_{jj',k}^{12}\right) = Var\left(X_{1jk} - X_{2j'k}\right) = 2\sigma_{R}^{2} + 2\sigma_{RC}^{2} + 2\sigma_{mR}^{2} + 2\sigma_{mC}^{2} + 2\sigma_{\varepsilon}^{2} \,. \quad (8)$$

To estimate these variances of differences, we need to estimate the variance components. From the three-way mixed effect ANOVA table (Table 1), we can relate the variance components with the mean squares. Therefore, the estimates of the variances of differences are

$$\hat{V}_{WR}^{ii} = 2MSE \,, \quad (9)$$

$$\hat{V}_{BR}^{ii} = \frac{2}{IJK}\left[J * MSR + J(K-1) * MSRC + J(I-1) * MSMR + (IJK - IJ - JK + J) * MSE\right], \quad (10)$$

$$\hat{V}_{WR}^{12} = \frac{2}{JK}\left[J * MSMR + K * MSMC + (JK - J - K) * MSE\right], \quad (11)$$

$$\hat{V}_{BR}^{12} = \frac{2}{IJK}\left[J * MSR + J(K-1) * MSRC + J(I-1) * MSMR + IK * MSMC + (IJK - IJ - IK - JK + J) * MSE\right]. \quad (12)$$

### *2.4 Simulation model*

To verify the estimates of limits of agreement for MRMC data, we need a simulation model. A popular MRMC simulation model is the MRMC ROC simulation model of Roe and Metz (Roe & Metz, 1997). Gallas et al. illustrated how to adapt the Roe and Metz model for an agreement study in their paper (Gallas, Anam, Chen, Wunderlich, & Zhang, 2016). After the first step of eliminating the subscript for the truth state, the adapted Roe and Metz model is the same as the three-way mixed effect ANOVA model as shown in equation (4). However, this model is not adequate for simulating MRMC agreement data. As shown in the previous subsection, when we calculate the variance of within-reader differences, $V_{WR}^{ii}$ and $V_{WR}^{12}$, the variance components for the reader and case random effect, $\sigma_R^2$ and $\sigma_C^2$, do not affect the result. This is because the reader and case effects cancel out when we calculate the difference in measurements. Also,

$$Var\left(X_{ijk}|R_j\right) = Var\left(X_{ij'k}|R_{j'}\right) = \sigma_C^2 + \sigma_{RC}^2 + \sigma_{mR}^2 + \sigma_{mC}^2 + 2\sigma_\varepsilon^2 \,, \quad (13)$$

which means that for any given reader, the variability of the measurements is the same. Therefore, the readers simulated from the Roe and Metz model fail to reflect any differences in reader variability when differences may be expected due to differences in a reader's background, training, and expertise.

Gallas et al. (Gallas, Anam, Chen, Wunderlich, & Zhang, 2016) proposed a hierarchical model to overcome the short-comings of the Roe and Metz model for simulating agreement data. Compared to the linear structure of the Roe and Metz model, the hierarchical structure avoids the cancelling of reader and case effects when computing the differences in measurements. However, there are two limitations of the hierarchical model. One is that the covariance among the measurements are highly dependent on the variance of the true value, instead of the parameters related to the reader and case effect. The other is that there is no explicit marginal distribution of the decision score. Even though the numerical integration can be applied to compute any metric aggregate over all the readers and cases, it will be more convenient if we have a model that has an explicit marginal distribution.

In this paper, we propose the following normal-inverse-gamma (Normal-IG) hierarchical simulation model:

$$X_{ijkl} = \mu + \tau_i + [RC]_{jk} + [\tau RCE]_{ijkl} \, , \tag{14}$$

where

$$[RC]_{jk} \mid R_j, \, C_k \sim N\left(C_k, R_j\right) \tag{15}$$

and

$$[\tau RCE]_{ijkl} \mid [\tau R]_{ij}, [\tau C]_{ik} \sim N\left([\tau C]_{ik}, [\tau R]_{ij}\right). \tag{16}$$

Also, $C_k \sim N(0, \sigma_C^2)$, $R_j \sim \text{InvG}(\alpha_R, \beta_R)$, $[\tau C]_{ik} \sim N(0, \sigma_{\tau C}^2)$, $[\tau R]_{ij} \sim \text{InvG}(\alpha_{\tau R}, \beta_{\tau R})$. Table 2 compares the model structure between the two hierarchical models. There are three key differences of the new proposed model compared to the previous hierarchical model. The first one is the position of the case effect in the conditional distribution of the interaction terms. Instead of contributing to the conditional variance of the interaction terms, in this model the case effect contributes to the conditional mean of the interaction terms. So $Cov\left(X_{ijkl}, X_{i'j'kl}\right) = \sigma_C^2$ is determined by the case related parameter, instead of the variance of the true quantitative value in the population. Since the normal distribution is a conjugate prior of a normal likelihood with unknown mean, the conditional distribution of the interaction term $[RC]_{jk}$ given reader effect $R_j$ is also normal , that is,

$$[RC]_{jk} \mid R_j \sim N(0, R_j + \sigma_C^2). \tag{17}$$

This also shows that for different readers, reader $j$ and $j'$, the conditional variance of the measurement score given $R_j$ and $R_{j'}$ will be different.

The second difference is the distribution for the reader effect. Here we use the inverse-Gamma distribution instead of the exponential distribution. The $\alpha_R$ and $\beta_R$ are shape and scale parameters of the reader effect. With larger $\alpha_R$, the variance of the reader will be smaller. Since the inverse-Gamma distribution is the conjugate prior of a normal likelihood with unknown variance, the interaction term $[RC]_{jk}$ conditional on case effect $C_k$is a scaled and shifted t-distributed with $2\alpha_R$ degrees of freedom. As the degree of freedom increases, the t-distribution becomes closer to the normal distribution. Therefore, with $\alpha_R$ and $\alpha_{\tau R}$ large enough, $X_{ijkl}$ is asymptotically normally distributed with mean

$$\mu + \tau_i \tag{18}$$

and variance

$$\frac{\beta_R}{\alpha_R - 1} + \sigma_C^2 + \frac{\beta_{\tau R}}{\alpha_{\tau R} - 1} + \sigma_{\tau C}^2 \,. \tag{19}$$

This gives us an approximate explicit marginal distribution for the measurement.

The third difference is the interaction term with modality and replicates. In this model, we eliminate the three-way interaction terms, reader-case-modality interaction and reader-case-replicate interaction, and only keep the two-way interaction of reader and case and the four-way interaction of all the four effects. In this way, the number of parameters in the models are controlled at a reasonable amount.

### *2.5 Derived means and variances of differences from the simulation model*

The true values for the WRBM and BRBM mean differences both equal the difference in modality effects. The true values for the variances of differences are affected by the parameters related to the reader and case distributions, which means that both reader and case variability contribute to the true value for the variances. The variances of the BRBM difference will be larger than the one for the WRBM difference, since it includes the variability among the different readers reading the same case. The detailed derivations are in the appendix.

$$\bar{D}_{WR}^{12} = \tau_i - \tau_{i'}, \tag{20}$$

$$V_{WR}^{12} = 2\sigma_{\tau C}^2 + \frac{2\beta_{\tau R}}{\alpha_{\tau R}-1}, \tag{21}$$

$$\bar{D}_{BR}^{12} = \tau_i - \tau_{i'}, \tag{22}$$

$$V_{BR}^{12} = 2\sigma_{\tau C}^2 + \frac{2\beta_{\tau R}}{\alpha_{\tau R}-1} + \frac{2\beta_R}{\alpha_R-1}. \tag{23}$$

***2.6 Verify the simulation is consistent with the derived true values***

By simulating individual WRBM and BRBM differences independently from the model, we can calculate Monte Carlo estimates of the variances of WRBM and BRBM differences and compare them to the true values derived from the model. There were two sets of parameters in this experiment. In both parameter sets, $\beta_R = \beta_{\tau R} = 1$, so that the reader variability is only affected by $\alpha_R$. One set of parameters fixed the case related parameters $\sigma_C^2 = \sigma_{\tau C}^2 = 1$ and allowed the reader related parameter $\alpha_R (= \alpha_{\tau R})$ to range from 2 to 20. For this set of parameters, the reader variability contributes 5% ~50% to the true value for variance of WRBM difference and 9.2% ~ 66.7% to the true value for variance of BRBM difference. A second set of parameters fixed the reader related parameters $\alpha_R = \alpha_{\tau R} = 10$ and allowed the case related parameter $\sigma_C^2 (= \sigma_{\tau C}^2)$ to range from 0.1 to 2, incrementing by 0.1. For this set of parameters, the case variability contributes 47.4% ~ 94.7% to the true value for variance of WRBM difference and 31% ~ 90% to the true value for variance of BRBM difference We simulated 100,000 trials with each trial having 4 measurements from 2 readers for a single case under 2 modalities. The Monte Carlo estimates of the variances are the sample variances of the 100,000 independent WRBM and BRBM differences. The relative bias between the Monte Carlo estimates $\tilde{V}_{WR}^{12}, \tilde{V}_{BR}^{12}$ and the derived true values $V_{WR}^{12}, V_{BR}^{12}$ are defined as

$$relative_bias\left(\tilde{V}_{WR}^{12}\right) = \frac{\tilde{V}_{WR}^{12} - V_{WR}^{12}}{V_{WR}^{12}}, \quad (24)$$

$$relative_bias\left(\tilde{V}_{BR}^{12}\right) = \frac{\tilde{V}_{BR}^{12} - V_{BR}^{12}}{V_{BR}^{12}}. \quad (25)$$

***2.7 Validate and characterize the MRMC limits of agreement estimates***

In this experiment, we compare the variance estimates of WRBM and BRBM differences by ANOVA to the derived true values. We denote the variance estimates of WRBM and BRBM differences for the trial $t$ by $\hat{V}_{WR,t}^{12}$ and $\hat{V}_{BR,t}^{12}$ and we assess the estimates in terms of relative bias and coefficient of variation:

$$relative_bias\left(\hat{V}_{WR}^{12}\right) = \frac{\frac{1}{T}\sum_t \hat{V}_{WR,t}^{12} - V_{WR}^{12}}{V_{WR}^{12}}, \quad (26)$$

$$CV\left(\hat{V}_{WR}^{12}\right) = \frac{SD\left(\hat{V}_{WR,t}^{12}\right)}{V_{WR}^{12}}. \quad (27)$$

where $SD()$ denotes the sample standard deviation across $T = 1000$ Monte Carlo trials.

We tested on the following three sets of parameters:

(1) Different number of readers $J = 3,4, \dots ,10, K = 100, \alpha_R = 6, \sigma_C^2 = 0.4$

(2) Different number of cases $K = 50,60, \ldots ,150, J = 5, \alpha_R = 6, \sigma_C^2 = 0.4$

(3) Different reader and case variabilities $(\alpha_R, \sigma_C^2) \in \{3,4,6,11,21\} \times \{0.1,0.2,0.4, 2/3, 1\}$, $J = 5, K = 100$

The range of $\alpha_R$ and $\sigma_C^2$ in the third set was selected based on the weights of parameters in true value $V_{BR}^{12}$. When $\beta_R = \beta_{\tau R} = 1, \alpha_R = \alpha_{\tau R}, \sigma_C^2 = \sigma_{\tau C}^2$,

$$V_{BR}^{12} = 2\sigma_C^2 + \frac{4}{\alpha_R - 1} \tag{28}$$

To make sure both $\alpha_R$ and $\sigma_C^2$ range from low variability to high variability and contribute to $V_{BRBM}$ at similar levels, we selected the parameters so that $4/(\alpha_R - 1)$ and $2\sigma_C^2$ take the values 0.2,0.4,0.8, $4/3$, 2. Hence, the true value of $V_{BR}^{12}$ ranged from 0.4 to 4, and when $\alpha_R = 6, \sigma_C^2 = 0.4$, the contributions to $V_{BR}^{12}$ from reader and case variability are both equal to 0.8.

## 3. Results

### *3.1 Verify the simulation is consistent with the derived true values*

In Figure 3, we show the relative bias observed between Monte Carlo estimates $\tilde{V}_{WR}^{12}, \tilde{V}_{BR}^{12}$ and the derived true value $V_{WR}^{12}, V_{BR}^{12}$ while varying the reader related parameter $\alpha_R$ and case related parameter $\sigma_C^2$. The dashed horizontal lines in each of the subplots denote the 0 bias between the Monte Carlo estimation and true values. As shown in the plots, the relative bias observed is distributed tightly around 0 and the absolute value of the relative bias is less than 1% for most of the cases. There is no linear trend with respect to the changes of the parameter settings, which indicates that the bias is independent of the magnitudes of the true value. These results are consistent with an estimator expected to be unbiased.

### *3.2 Validate and characterize the MRMC limits of agreement estimates*

Figure 4 and Figure 5 present the relative bias and CV of the variance estimates of BRBM differences changing over different sets of parameters. The plots for the WRBM differences share the similar pattern, so only the ones for the BRBM difference are shown. In Figure 4, plots (a) and (c) show the results when the number of readers ranges from 3 to 10 and the number of cases is 100, while (b) and (d) shows the results when the number of cases changes from 50 to 150 and the number of readers is 5. In both settings, the relative bias observed is small. This is comparable to the relative bias in the previous study. When we compare the CVs for the two sets of simulation parameters, the CVs of the variance estimates of BRBM in both plots (c) and (d) decreases smoothly as the number of readers and cases increases. This is because as we increase the size of the MRMC study, the variance estimate is more precise for each simulation study. Therefore, the variation across all the simulation studies will decrease.

Figure 5 describes how the relative bias and CV of the variance estimates for BRBM differences change over the reader and case related parameters, $\alpha_R$ and $\sigma_C^2$. There are 5 readers and 100 cases in each MRMC study. From Figure 5(a) we can see that most of the points lie within the range of ±1%. The line with solid dot ( $\alpha_R = 3$ ) has the largest fluctuation compared to other lines. This is the setting that involves the largest reader variability, so it is expected that the variance of differences, especially for the between-reader difference, will be less stable than the results for other parameter settings. This can also be confirmed in Figure 5(b); the line for $\alpha_R = 3$ (line with solid dot) has the highest CV for each $\sigma_C^2$. Generally, as $\alpha_R$ increases, the CV drops, since the reader variability in the study decreases. As we increase the case related parameter $\sigma_C^2$, the case variability increases and it will dominate the variability from the readers, since the number of cases is much larger than that for the readers. Therefore, the lines in Figure 5(b) will be closer to each other when the $\sigma_C^2$ increases from 0.1 to 1.

## 4. Discussions and Conclusion

In this work, we focused on the analysis of quantitative MRMC data for the clinical task with no underlying binary ground truth. We introduced four different types of LOA for MRMC agreement studies comparing two modalities, and we proposed to apply an ANOVA model for estimating the LOA. For the clinical task that has a binary ground truth or ordinal reference standard, the MRMC ROC analysis can generate reader-averaged AUC and its uncertainty for evaluating the reader performance. Different ANOVA models have been widely applied to do MRMC ROC analysis. For example, Obuchowski (Obuchowski N. A., 1995) used a two-way mixed effect ANOVA model with correlated error terms to analyze AUC values for different reader-modality combinations, and Dorfman et al. (Dorfman, Berbaum, & Metz, 1992) used a three-way mixed effect model to analyze the jackknife pseudovalues for each case. Though we use the same three-way mixed effect model structure in this paper, the model is directly applied to the quantitative measurements from each reader for each case and there is no need for computing pseudovalues. If an MRMC study has neither binary ground truth nor quantitative reference standard and the data generated from the study is binary outcome data (agree or disagree with reference), then it is called MRMC study with binary agreement data. The interested reader is referred to Chen et al. (Chen, Wunderlich, Petrick, & Gallas, 2014) for its analysis and sizing.

Tcheuko et al. (Tcheuko, Gallas, & Samuelson, 2016) proved that the two-way random effect ANOVA model can be applied to estimate the variance of a two-sample U-statistic of order (1,1), and the estimate is equal to the U-statistics estimate of the variance. If we regard the WRBM differences, $D_{j,k}^{12} = X_{1jk} - X_{2jk}$, as the U-statistics kernel, the estimate of mean differences $\widehat{\bar{D}}_{WR}^{12}$ is also a U-statistic of order (1,1). Then, we can use both the U-statistics estimation method and estimates from a two-way random effect ANOVA model to estimate the variance of WRBM differences. The results are equivalent to the variance estimate from the three-way mixed effect ANOVA method as we presented above. When it comes to the BRBM differences, $D_{jj',k}^{12} = X_{1jk} - X_{2j'k}$, the estimate of the mean value over all pairs of readers and all cases, $\widehat{\bar{D}}_{BR}^{12}$, is still a U-statistic. However, it is not of order (1,1) but of order (2,1) as it involves two readers and one case for generating the between-reader difference measurements. Hence, the two-way ANOVA method is not suitable for estimating variance of WRBM differences. In general, a three-way mixed effect ANOVA method is able to estimate the variances of all four types of MRMC differences, whereas a two-way random effect ANOVA method can only estimate the variances of the within-reader differences.

To analyze quantitative MRMC data with an ANOVA model, we need the study to be fully crossed and the measurements should be paired for the test modality and reference modality. In some situations, the reference standards and the testing measurements are generated from different groups of readers, or a split-plot study is applied to reduce the work load for each reader (Obuchowski, Gallas, & Hillis, 2012). Then, the current analysis may not work. Therefore, future work should extend this work to the MRMC study that is not fully crossed or not paired.

Another possible direction to extend the current work is to estimate the precision of the WRBM and BRBM LOA. Bland and Altman proposed a method to estimate the confidence limits for the upper and lower limits of agreement, based on the independent and normal assumption for the individual differences (Bland & Altman, 1999). The difference measurements in an MRMC study are correlated due to the measurements are from the same readers or the measurements are for the same cases. So the calculation of the confidence limits for the LOA in an MRMC study will be more complicated.

Finally, the current analysis is for characterizing the data shown in a Bland-Altman plot for an MRMC study. A Bland-Altman plot helps visualize the distribution of the difference measurements as a function of the magnitude of the measurements. However, there is no hypothesis testing method to help us decide whether there are statistically significant differences between the two modalities. Thus, it is important that future work extend the current analysis to hypothesis testing and sizing methods for the MRMC agreement study.

In conclusion, we proposed a simulation model for generating quantitative MRMC data. Compared to the Roe & Metz model that is commonly used for simulating MRMC ROC data, the data generated from the new simulation model can reflect reader variability in the conditional variance for different readers. With the case component contributing to the conditional mean and the reader component contributing to the condition variance in the hierarchical model structure, we assume the reader components do not affect the mean value of the measurement but influnce the correlation structure of the measurements. This also facilitates simulating measurements that are not normally distributed. We can just change the distributions of the case effects, which, will not affect the correlation structure of the measurements This also reduces the number of parameters needed in the simulation model. From the simulation results, we can conclude that our estimation results for the variances of WRBM and BRBM differences are unbiased and the uncertainty of the estimation drops as the number of readers and cases increases and rises as the value of true variance increases.

The code used for the simulation study and ANOVA estimates in this paper are available at ANOVA.MRMC.LOA (github.com)

## APPENDIX: DERIVATION OF MEANS AND VARIANCES OF DIFFERENCES FROM THE SIMULATION MODEL

The true values for the means of WRBM and BRBM differences can be derived by applying the law of total expectation.

$$\bar{D}_{WR}^{12} = E(X_{ijkl} - X_{i'jkl}) = E(\tau_i - \tau_{i'} + [\tau RCE]_{ijkl} - [\tau RCE]_{i'jkl})$$

$$= \tau_i - \tau_{i'} + E\left(E\left([\tau RCE]_{ijkl} - [\tau RCE]_{i'jkl} | [\tau R]_{ij}, [\tau C]_{ik}, [\tau R]_{i'j}, [\tau C]_{i'k}\right)\right)$$

$$\bar{D}_{BR}^{12} = E(X_{ijkl} - X_{i'j'kl}) = E(\tau_i - \tau_{i'} + [RC]_{jk} - [RC]_{j'k} + [\tau RCE]_{ijkl} - [\tau RCE]_{i'j'kl})$$

$$= \tau_i - \tau_{i'} + E\left(E([RC]_{jk} - [RC]_{j'k} | R_j, R_{j'}, C_k)\right)$$

$$+ E\left(E\left([\tau RCE]_{ijkl} - [\tau RCE]_{i'j'kl} | [\tau R]_{ij}, [\tau C]_{ik}, [\tau R]_{i'j'}, [\tau C]_{i'k}\right)\right)$$

From Table 2 we know the conditional mean of the interaction terms given the reader and case effects. That is, $E([RC]_{jk} | R_j, C_k) = C_k$ and $E([\tau RCE]_{ijkl} | [\tau R]_{ij}, [\tau C]_{ik}) = [\tau C]_{ik}$. Thus, the expected WRBM and BRBM differences can be simplified as:

$$\bar{D}_{WR}^{12} = \tau_i - \tau_{i'} + E([\tau C]_{ik} - [\tau C]_{i'k}),$$

$$\bar{D}_{BR}^{12} = \tau_i - \tau_{i'} + E(C_k - C_k) + E([\tau C]_{ik} - [\tau C]_{i'k}).$$

From the distribution of the case effect shown in Table 2, we know that both $C_k$ and $[\tau C]_{ik}$ are normally distributed with mean 0. Therefore,

$$\bar{D}_{WR}^{12} = \bar{D}_{BR}^{12} = \tau_i - \tau_{i'}.$$

Similarly, the true values for the variances of WRBM and BRBM differences can be derived by applying the law of total variance.

$$V_{WR}^{12} = Var(X_{ijkl} - X_{i'jkl}) = Var(\tau_i - \tau_{i'} + [\tau RCE]_{ijkl} - [\tau RCE]_{i'jkl})$$

$$= Var\left(E\left([\tau RCE]_{ijkl} - [\tau RCE]_{i'jkl} | [\tau R]_{ij}, [\tau C]_{ik}, [\tau R]_{i'j}, [\tau C]_{i'k}\right)\right)$$

$$+ E\left(Var\left([\tau RCE]_{ijkl} - [\tau RCE]_{i'jkl} | [\tau R]_{ij}, [\tau C]_{ik}, [\tau R]_{i'j}, [\tau C]_{i'k}\right)\right)$$

$$V_{BR}^{12} = Var\left(X_{ijkl} - X_{i'j'kl}\right)$$

$$= Var\left(\tau_i - \tau_{i'} + [RC]_{jk} - [RC]_{j'k} + [\tau RCE]_{ijkl} - [\tau RCE]_{i'j'kl}\right)$$

$$= Var\left(E\left([RC]_{jk} - [RC]_{j'k}|R_j, R_{j'}, C_k\right)\right) + E\left(Var\left([RC]_{jk} - [RC]_{j'k}|R_j, R_{j'}, C_k\right)\right)$$

$$+ Var\left(E\left([\tau RCE]_{ijkl} - [\tau RCE]_{i'j'kl}|[\tau R]_{ij}, [\tau C]_{ik}, [\tau R]_{i'j'}, [\tau C]_{i'k}\right)\right)$$

$$+ E\left(Var\left([\tau RCE]_{ijkl} - [\tau RCE]_{i'jkl}|[\tau R]_{ij}, [\tau C]_{ik}, [\tau R]_{i'j'}, [\tau C]_{i'k}\right)\right)$$

From Table 2 we know the conditional mean and variance of the interaction terms given the reader and case effects. That is, $E\left([RC]_{jk}|R_j, C_k\right) = C_k$, $Var\left([RC]_{jk}|R_j, C_k\right) = R_j$, $E\left([\tau RCE]_{ijkl}|[\tau R]_{ij}, [\tau C]_{ik}\right) = [\tau C]_{ik}$ $Var\left([\tau RCE]_{ijkl}|[\tau R]_{ij}, [\tau C]_{ik}\right) = [\tau R]_{ij}$. Thus, the variances of WRBM and BRBM differences can be simplified as:

$$V_{WR}^{12} = Var([\tau C]_{ik} - [\tau C]_{i'k}) + E\left([\tau R]_{ij} + [\tau R]_{i'j}\right),$$

$$V_{BR}^{12} = Var(C_k - C_k) + E\left(R_j + R_{j'}\right) + Var([\tau C]_{ik} - [\tau C]_{i'k}) + E\left([\tau R]_{ij} + [\tau R]_{i'j'}\right).$$

Again, from the distributions of the reader and case effects in Table 2, we know that $Var([\tau C]_{ik}) = \sigma_{\tau C}^2$, $E\left(R_j\right) = \frac{\beta_R}{\alpha_R - 1}$, and $E\left([\tau R]_{ij}\right) = \frac{\beta_{\tau R}}{\alpha_{\tau R} - 1}$. Also, the terms with different subscripts are independently and identically distributed.

$$V_{WR}^{12} = 2\sigma_{\tau C}^2 + \frac{2\beta_{\tau R}}{\alpha_{\tau R} - 1}$$

$$V_{BR}^{12} = \frac{2\beta_R}{\alpha_R - 1} + 2\sigma_{\tau C}^2 + \frac{2\beta_{\tau R}}{\alpha_{\tau R} - 1}$$

Table 1 Three-way mixed effect ANOVA table

| Source | DF | Sum of Square (SS) | Mean Square (MS) | E(MS) |
|---|---|---|---|---|
| **Modality** | $I-1$ | $SSM = JK\sum_i(\overline{X_{i\cdot\cdot}} - \overline{X_{\cdots}})^2$ | $MSM = \frac{SSM}{I-1}$ | $\sigma_\varepsilon^2 + K\sigma_{mR}^2 + J\sigma_{mC}^2 + \frac{JK}{I-1}\sum_i m_i^2$ |
| **Reader** | $J-1$ | $SSR = IK\sum_j(\overline{X_{\cdot j\cdot}} - \overline{X_{\cdots}})^2$ | $MSR = \frac{SSR}{J-1}$ | $\sigma_\varepsilon^2 + I\sigma_{RC}^2 + K\sigma_{mR}^2 + IK\sigma_R^2$ |
| **Case** | $K-1$ | $SSC = IJ\sum_i(\overline{X_{\cdot\cdot k}} - \overline{X_{\cdots}})^2$ | $MSC = \frac{SSC}{K-1}$ | $\sigma_\varepsilon^2 + I\sigma_{RC}^2 + J\sigma_{mC}^2 + IJ\sigma_C^2$ |
| **Reader: Case** | $(J-1)(K-1)$ | $SSRC = I\sum_j\sum_k(\overline{X_{\cdot jk}} - \overline{X_{\cdot j\cdot}} - \overline{X_{\cdot\cdot k}} + \overline{X_{\cdots}})^2$ | $MSRC = \frac{SSRC}{(J-1)(K-1)}$ | $\sigma_\varepsilon^2 + I\sigma_{RC}^2$ |
| **Reader: Modality** | $(I-1)(J-1)$ | $SSMR = K\sum_i\sum_j(\overline{X_{ij\cdot}} - \overline{X_{i\cdot\cdot}} - \overline{X_{\cdot j\cdot}} + \overline{X_{\cdots}})^2$ | $MSMR = \frac{SSMR}{(I-1)(J-1)}$ | $\sigma_\varepsilon^2 + K\sigma_{mR}^2$ |
| **Case: Modality** | $(I-1)(K-1)$ | $SSMC = J\sum_i\sum_k(\overline{X_{i\cdot k}} - \overline{X_{i\cdot\cdot}} - \overline{X_{\cdot\cdot k}} + \overline{X_{\cdots}})^2$ | $MSMC = \frac{SSMC}{(I-1)(K-1)}$ | $\sigma_\varepsilon^2 + J\sigma_{mC}^2$ |
| **Error** | $df_{Error}$ * | $SSE = SST - SSM - SSR - SSC - SSRC - SSMR - SSMC$ | $MSE = \frac{SSE}{df_{Error}}$ | $\sigma_\varepsilon^2$ |
| **Total** | $IJK-1$ | $SST = \sum_i\sum_j\sum_k(X_{ijk} - \overline{X_{\cdots}})^2$ | | |

* $df_{Error} = IJK - IJ - JK - IK + I + J + K - 1$

Table 2 Compare the model structure of the two hierarchical models.

| Models | Interaction Term | Conditional Distribution of the Interaction Term Given Reader and Case Effects | | Case Effect Distribution | Reader Effect Distribution |
|---|---|---|---|---|---|
| | | Mean | Variance | | |
| **Normal-IG hierarchical model** | $[RC]_{jk}$ | $C_k$ | $R_j$ | $N(0,\sigma_C^2)$ | $\text{InvG}(\alpha_R,\beta_R)$ |
| | $[\tau RCE]_{ijkl}$ | $[\tau C]_{ik}$ | $[\tau R]_{ij}$ | $N(0,\sigma_{\tau C}^2)$ | $\text{InvG}(\alpha_{\tau R},\beta_{\tau R})$ |
| **Gallas hierarchical model** | $[RC]_{jk}$ | 0 | $(R_j+C_k)^2$ | $Exp(1/\mu_C)$ | $Exp(1/\mu_R)$ |
| | $[\tau RC]_{ijk}$ | 0 | $([\tau R]_{ij}+[\tau C]_{ik})^2$ | $Exp(1/\mu_{\tau C})$ | $Exp(1/\mu_{\tau R})$ |
| | $[RCE]_{jkl}$ | 0 | $([RE]_{jl}+[CE]_{kl})^2$ | $Exp(1/\mu_{CE})$ | $Exp(1/\mu_{RE})$ |
| | $[\tau RCE]_{ijkl}$ | 0 | $([\tau RE]_{ijl}+[\tau CE]_{ikl})^2$ | $Exp(1/\mu_{\tau CE})$ | $Exp(1/\mu_{\tau RE})$ |

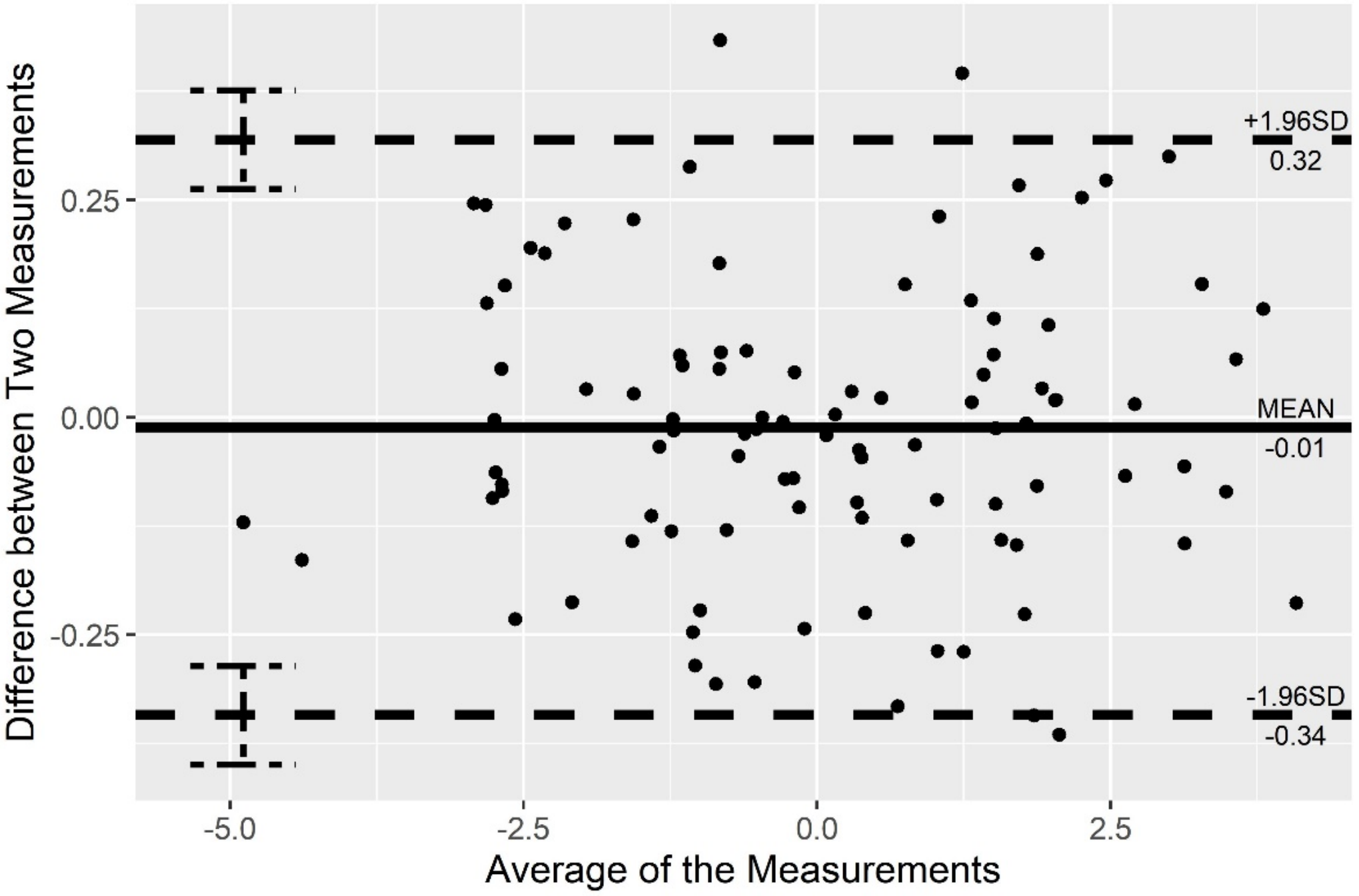

Figure 1 An example Bland-Altman plot for independent differences. The two groups of measurements are simulated from the distributions: $Y_{1k}|C_k = x \sim N(2x, 0.1), Y_{2k}|C_k = x \sim N(2x, 0.11)$, and $C_k \sim N(0,1)$. We simulate 100 independent cases ($C_k$). For each case, one $Y_{1k}$ and one $Y_{2k}$ are simulated. They are shown as the dots in the plot. The solid line with annotation "MEAN" shows the mean difference of the measurements. The two dashed lines with the annotations "+1.96SD" and "-1.96SD" represent the upper and lower bound of 95% limits of agreement. The error bars on the dashed lines represent the confidence intervals for the limits of agreement.

| Within-Modality | | Modality $i$ | | | |
|---|---|---|---|---|---|
| | | Reader1 | Reader2 | … | Reader$J$ |
| Modality $i$ | Reader1 | $\bar{D}_{11}^{ii}$ | $\bar{D}_{12}^{ii}$ | … | $\bar{D}_{1J}^{ii}$ |
| | Reader2 | $\bar{D}_{21}^{ii}$ | $\bar{D}_{22}^{ii}$ | $\ddots$ | $\vdots$ |
| | $\vdots$ | $\vdots$ | $\ddots$ | $\ddots$ | $\bar{D}_{(J-1)J}^{ii}$ |
| | Reader $J$ | $\bar{D}_{J1}^{ii}$ | … | $\bar{D}_{J(J-1)}^{ii}$ | $\bar{D}_{JJ}^{ii}$ |

| Between-Modality | | Reference Modality $i = 2$ | | | |
|---|---|---|---|---|---|
| | | Reader1 | Reader2 | … | Reader$J$ |
| Test Modality $i = 1$ | Reader1 | $\bar{D}_{11}^{12}$ | $\bar{D}_{12}^{12}$ | … | $\bar{D}_{1J}^{12}$ |
| | Reader2 | $\bar{D}_{21}^{12}$ | $\bar{D}_{22}^{12}$ | $\ddots$ | $\vdots$ |
| | $\vdots$ | $\vdots$ | $\ddots$ | $\ddots$ | $\bar{D}_{(J-1)J}^{12}$ |
| | Reader $J$ | $\bar{D}_{J1}^{12}$ | … | $\bar{D}_{J(J-1)}^{12}$ | $\bar{D}_{JJ}^{12}$ |

Figure 2 Case-averaged difference between measurements within- or between- readers and within- or between- modalities. The within-reader differences are on the diagonals and the between-reader differences are on the off-diagonals. The within-reader differences require each reader to read the cases twice, replicate data.

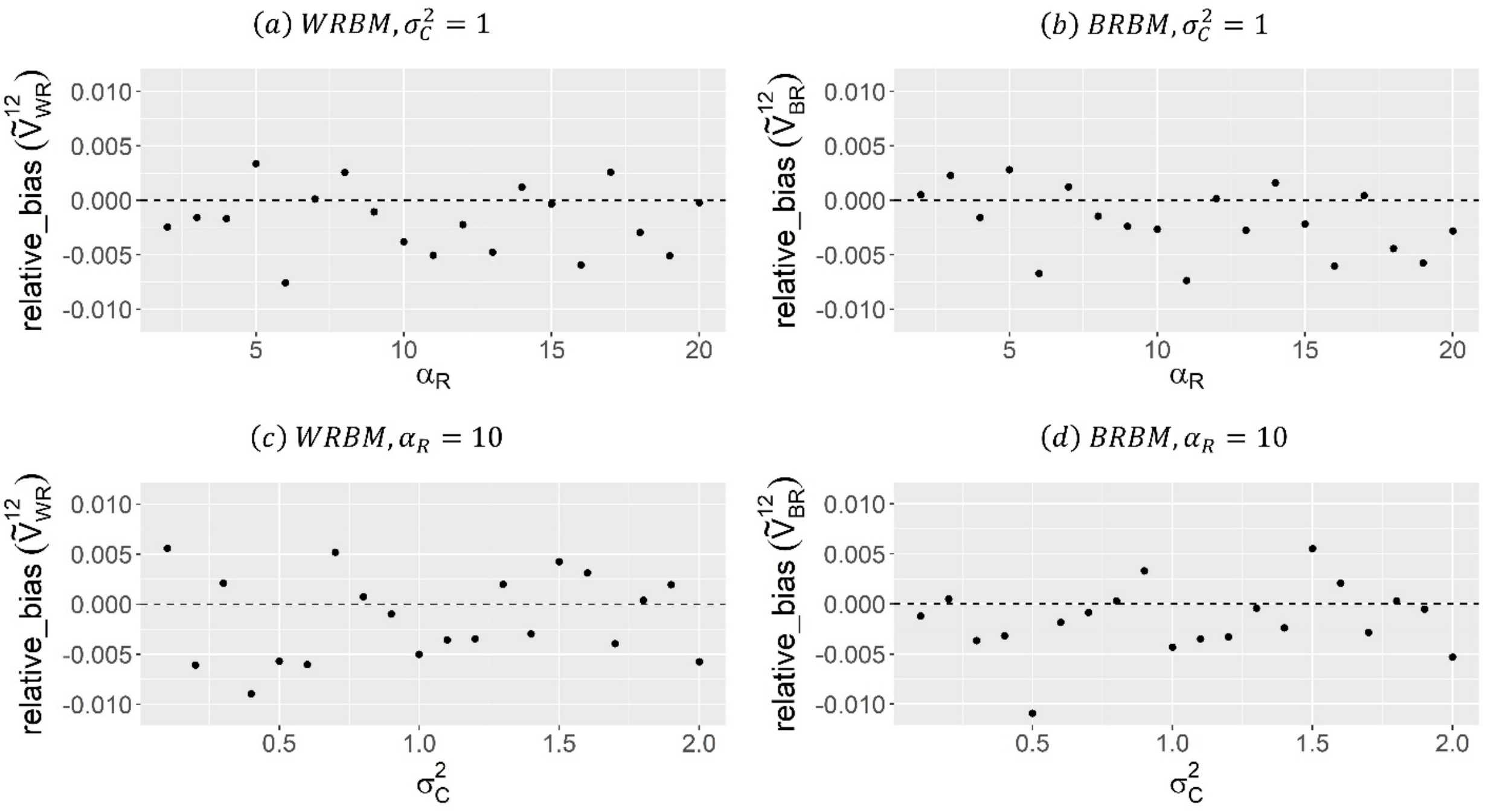

Figure 3 Relative bias between the Monte Carlo estimates ($T = 100{,}000$) and the derived true values while varying $\alpha_R$ and $\sigma_C^2$

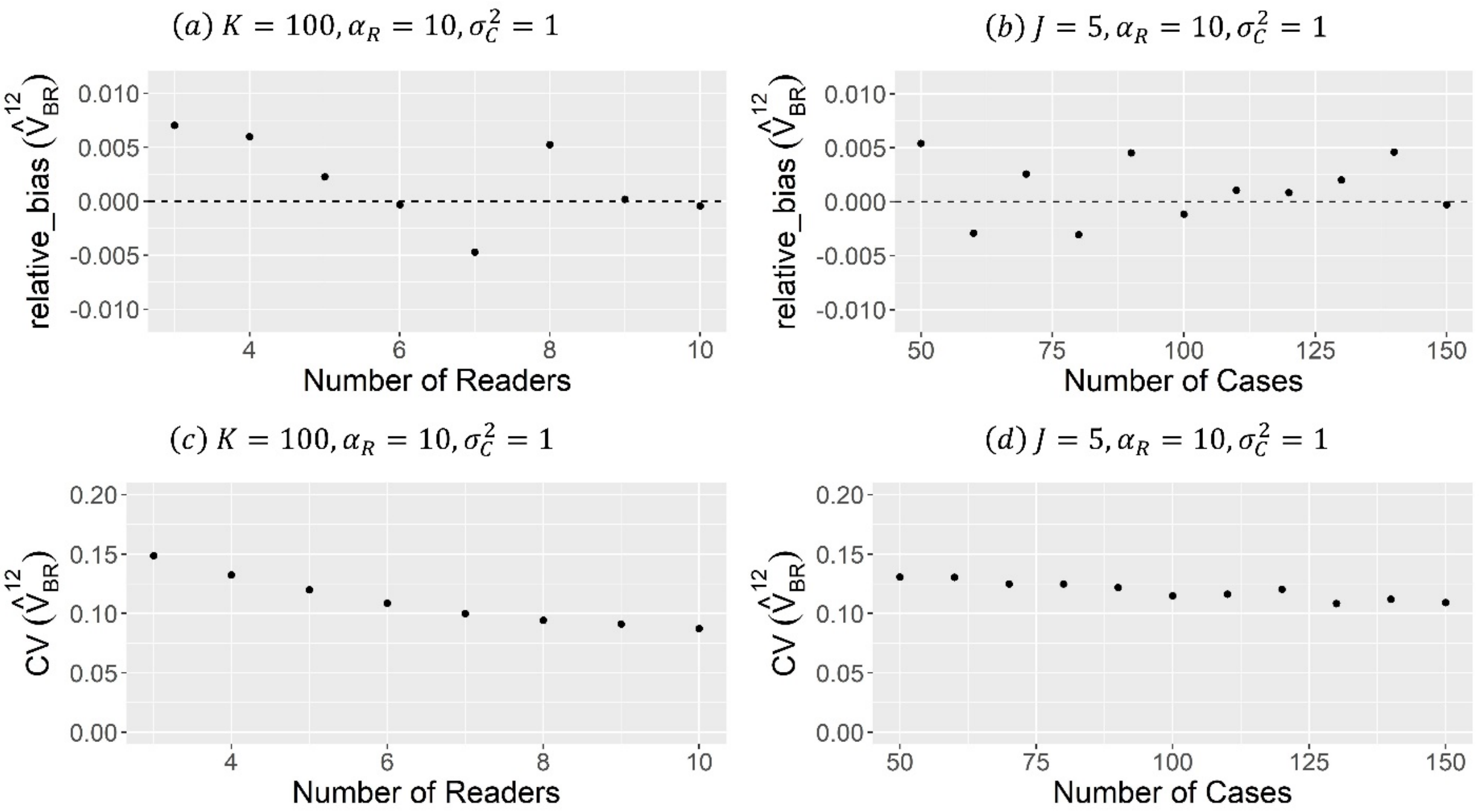

Figure 4 Relative bias and CV of variance estimates for BRBM differences changing over the size of the study

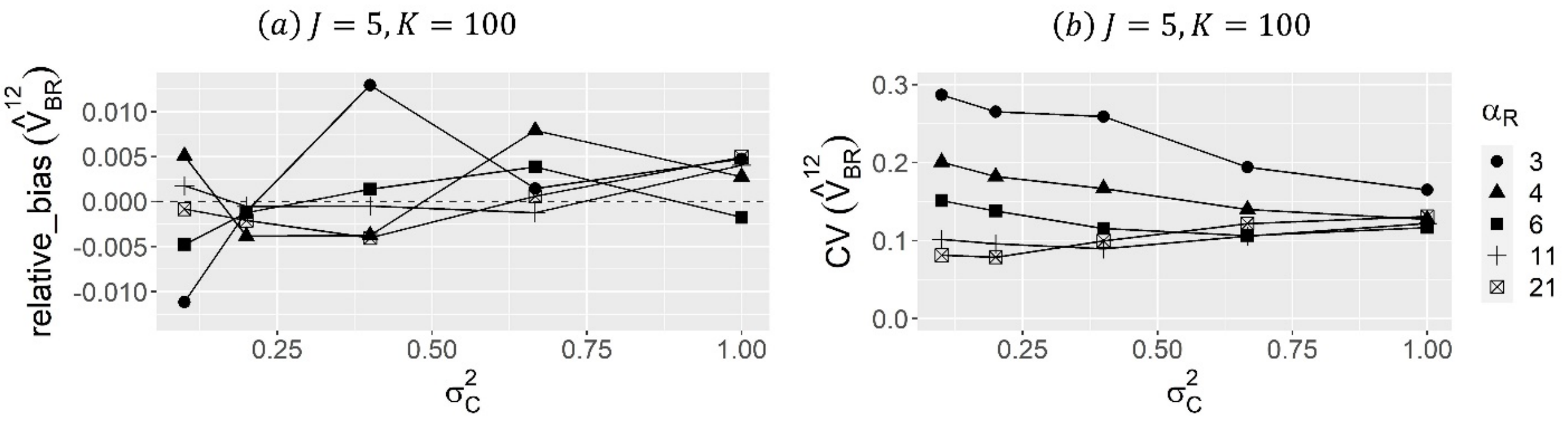

Figure 5 Relative bias and CV of variance estimates for BRBM difference with varying $\alpha_R$ and $\sigma_C^2$

Figure 1 An example Bland-Altman plot for independent differences. The two groups of measurements are simulated from the distributions: $YY_{1k}|C_k = x \sim N(2x, 0.1), Y_{2k}|C_k = x \sim N(2x, 0.11)$, and $C_k \sim N(0,1)$. We simulate 100 independent cases ($C_k$). For each case, one $Y_{1k}$ and one $Y_{2k}$ are simulated. They are shown as the dots in the plot. The solid line with annotation "MEAN" shows the mean difference of the measurements. The two dashed lines with the annotations "+1.96SD" and "-1.96SD" represent the upper and lower bound of 95% limits of agreement. The error bars on the dashed lines represent the confidence intervals for the limits of agreement.

Figure 2 Case-averaged difference between measurements within- or between- readers and within- or between- modalities. The within-reader differences are on the diagonals and the between-reader differences are on the off-diagonals. The within-reader differences require each reader to read the cases twice, replicate data.

Figure 3 Relative bias between the Monte Carlo estimates ($T = 100{,}000$) and the derived true values while varying $\alpha_R$ and $\sigma_C^2$

Figure 4 Relative bias and CV of variance estimates for BRBM differences changing over the size of the study

Figure 5 Relative bias and CV of variance estimates for BRBM difference with varying $\alpha_R$ and $\sigma_C^2$